\documentclass[aps,prl,twocolumn,showpacs,groupaddresses,superscriptaddress]{revtex4}
\usepackage{bbm,amssymb,amsmath,amsbsy,times,pifont,pxfonts}
\usepackage[dvips]{graphicx}
\usepackage{hyperref}
\usepackage[latin1]{inputenc}
\usepackage{subfigure}
\usepackage{fancyhdr,makeidx}
\usepackage{color}

\newcommand{\id}{\mathbbm{1}}

\newcommand{\gr}[1]{\boldsymbol{#1}}
\newcommand{\be}{\begin{equation}}
\newcommand{\ee}{\end{equation}}
\newcommand{\bea}{\begin{eqnarray}}
\newcommand{\eea}{\end{eqnarray}}
\newcommand{\ket}[1]{|#1\rangle}
\newcommand{\bra}[1]{\langle#1|}

\newcommand{\eq}[1]{Eq.~(\ref{#1})}

\begin{document}

\title{Noisy Quantum Cellular Automata for Quantum vs Classical Excitation Transfer}

\author{Michele Avalle and Alessio Serafini}
\affiliation{Department of Physics \& Astronomy, University College London, 
Gower Street, London WC1E 6BT, United Kingdom}
\email{michele.avalle.11@ucl.ac.uk}

\begin{abstract}

We introduce a class of noisy quantum cellular automata on a qubit lattice 
that includes all classical Markov chains, as well as maps where quantum coherence between sites is allowed to build up over time.
We apply such a construction to the problem of excitation transfer through 1-$d$ lattices, 
and compare the performance of classical and quantum dynamics with equal local transition probabilities.
Our discrete approach has the merits of stripping down the complications of the open system dynamics, 
of clearly isolating coherent effects, of allowing for an exact treatment of conditional dynamics,
all while capturing a rich variety of dynamical behaviours.

\end{abstract}


\maketitle

Quantifying the extent to which quantum coherence enhances the performance 
of antennae or communication systems is a timely \cite{Mohseni,Engel,lee,Hayes,Collini,Science,Mohseni2,rebentrost,rebentrost2,Plenio2,caruso,muelken,Olaya,
francesca,olayaview,Aspuru,Vedral2,vibquabio,Mercer}, yet often controversial, subject. 
Typically, a classical system is compared to the analogous, quantised model \cite{muelken,Olaya,ishizaki,berkelbach,wu},
although such a correspondence is not straightforward when open, dissipative systems are considered,
as it should be in most cases of interest. 

In this manuscript, we introduce a strictly local model of energy transfer via a noisy quantum cellular automaton  
construction \cite{Schumacher,ArrighiQCA} on a qubit lattice. 
Tuning one real parameter of such a model will allow us to range 
from a classical Markov chain, where quantum coherence is systematically suppressed at each time-step of the automaton, to dynamics where quantum coherence is allowed to build up over time, while keeping, by construction, 
the local transition probabilities constant. Thus, a ``fair'' comparison between classical and quantum energy transfer may be 
carried out, where the effect of quantum interference is singled out with no ambiguity. 
Our model, restricted to the first excitation subspace, can be studied exactly for very large systems, also including conditional 
dynamics due to measurements (which will model the absorption of the excitation at the end of the energy transfer process).

The plan of the paper is as follows.
We shall first consider the problem of constructing a class of one-qubit completely positive (CP) maps that, in a certain limit, reproduce all classical Markov transition matrices on dichotomic probability distributions. We will then apply our construction to the first excitation subspace of a partitioned quantum cellular automaton structure 
(where one-qubit maps will be applied to the two-dimensional space spanned by excitations at neighbouring sites), 
obtaining a global dynamics on a lattice which is capable of describing excitation transfer. 
We will then present a study of the performance of classical versus quantum maps, showing by how much and under what 
conditions does quantum coherence improve the probability of excitation transfer through the lattice. Finally, we shall draw some conclusions 
and discuss outlook of this work.

\noindent {\em CP-map representation of classical stochastic maps.} The most general classical stochastic 
map on dichotomic probability distributions, represented by vectors $\gr{v}_m=(m,1-m)^{\sf T}$ for $0\le m \le1$, is represented by 
a stochastic matrix of the form
\be
T_{p,q} = \left(\begin{array}{cc}
1-p & q \\
p & 1-q 
\end{array}\right) \quad {\rm with} \quad 0 \le p,q \le 1 \, . \label{stocco}
\ee
Let us remind the reader that the matrix (and map) $T_{p,q}$ is referred to as doubly stochastic if $p=q$. 

Any classical system may be thought of as a limiting instance of a quantum system.
Hence, the action of a generic $T_{p,q}$ on dichotomic probability vectors may be obtained by considering 
the action of a specific class ${\mathcal C}$ of completely positive (CP)-maps on ``classical'' states, by which we mean 
density operators that are diagonal in a certain basis of the Hilbert space. 
Given the privileged classical basis, the class ${\mathcal C}$ of ``classical'' CP-maps is comprised of all the CP-maps that send diagonal density matrices into diagonal density matrices. 
Here, we give a simple construction which allows one to reproduce any possible two-dimensional stochastic matrix 
by considering a subset of ${\mathcal C}$ acting on a one-qubit system.

Diagonal density matrices may be trivially bijectively mapped into dichotomic probability distributions 
as per ${G}:\varrho_m={\rm diag}(m,1-m)\mapsto\gr{v}_m=(m,1-m)^{\sf T}$, where we denoted such a bijection by $G$.  
We will conventionally refer to the parameter $m$ as the probability of populating the excited state of the qubit, or `excitation probability'.
In practice, the classical basis will be dictated by decoherence processes, as we will indicate later on. 
We shall refer to the map $\Phi_{\xi}$, with Kraus operators
\be
K_0=\sqrt{1-\xi}\id,\;
K_1=\sqrt{\xi}(\id+\sigma_z)/2 , \;
K_2=\sqrt{\xi}(\sigma_z-\id)/2 \, , \label{dephase}
\ee
with $0\le\xi\le1$, as the dephasing map ($\sigma_j$ for $j=x,y,z$ stand for the Pauli matrices). 
The CP-map $\Phi_{1}$, whose effect is setting to zero the off-diagonal elements while leaving the diagonal 
ones unchanged, will be called `complete', or `total' dephasing. 
Let us also define the amplitude damping channel $\Xi_{\eta}$, with 
Kraus operators 
\be
L_{0,\eta} = (\id+\sigma_z)/2 + \sqrt{1-\eta}(\id-\sigma_z)/2 , \; L_{1,\eta}=\sqrt{\eta}(\sigma_x+i\sigma_y)/2 \, , \label{ampli}
\ee
with $0\le\eta\le 1$. For future convenience, let us extend the definition of $\Xi_{\eta}$ to negative $\eta$ ($-1\le\eta\le0$), 
as the `swapped' amplitude damping channel, 
with Kraus operators $\sigma_x L_{0,|\eta|}\sigma_x$ and $\sigma_x L_{1,|\eta|} \sigma_x$. 

Let us now proceed by showing two simple statements concerning the relationship between classical stochastic maps and single qubit 
dynamics.\smallskip

\noindent  {\em \textbf{Proposition 1.} Any two-dimensional doubly stochastic map $T_{p,p}$ may be represented on diagonal density matrices by the action of a unitary map followed by complete dephasing.} \smallskip

\noindent {\em Proof.} Let $U_{\theta,\gr{\varphi}}$ be a generic $2\times2$ unitary parametrised as 
\be
U_{\theta,\gr{\varphi}} = \left(\begin{array}{cc}
\cos\theta & \sin\theta {\rm e}^{i\varphi_2} \\
-\sin\theta{\rm e}^{i\varphi_1} & \cos\theta {\rm e}^{i(\varphi_1+\varphi_2)}
\end{array}\right) \, ,
\ee
with $0\le\theta\le\pi$, $\gr{\varphi}=(\varphi_1,\varphi_2)$ and $0\le\varphi_1,\varphi_2\le2\pi$.
Then it is immediate to see that the action of $U_{\theta,\gr{\varphi}}$ on a diagonal density matrix, with excitation probability $m$, 
followed by total dephasing, is a diagonal density matrix with excitation probability $\sin(\theta)^2+\cos(2\theta)m$. 
This is analogous to the action of the stochastic map $T_{p,p}$ of \eq{stocco} on the probability vector $\gr{v}_m=(m,1-m)^{\sf T}$, 
upon identifying $p=\sin(\theta)^2$. 
In formulae:
\be
G\left(\Phi_1\left(U_{\theta,\gr{\varphi}} \varrho_m U_{\theta,\gr{\varphi}}^{\dag} \right)\right) = 
T_{\sin^2\theta,\sin^2\theta} \gr{v}_m \; .
\ee
It is hence evident that a proper choice of $\theta$ allows one to 
reproduce any doubly stochastic map. $\square$ \smallskip

\noindent {\em \textbf{Proposition 2.} Any two-dimensional stochastic map $T_{p,q}$ 
may be represented on diagonal density matrices by the action of a completely dephased unitary map, 
followed by an amplitude damping channel.} \smallskip

\noindent{\em Proof.} As we saw above, the action of a completely dephased unitary $U_{\theta,\gr{\varphi}}$ on a 
diagonal density matrix $\varrho_m$ yields the diagonal state $\varrho_{(1-c)/2+cm}$ with excitation probability 
$(1-c)/2+cm$ , where we shortened the notation by setting $c=\cos(2\theta)$. 
The action of a (direct or swapped, as indicated by the sign of $\eta$) amplitude damping channel $\Xi_{\eta}$ 
on the state $\varrho_{(1-c)/2+cm}$ leads to another diagonal state $\varrho_{m'}$ with excitation probability
\be
m' = c(1-|\eta|)m + \frac{1+|\eta| c -c +\eta}{2} \; . \label{muno1}
\ee
On the other hand, the action of $T_{p,q}$ on the vector $\gr{v}_m$ gives a vector $\gr{v}_{m'}$, with 
\be
m' = (1-p-q)m+q \; . \label{muno2}
\ee
By comparing Eqs.~(\ref{muno1}) and (\ref{muno2}), one obtains 
\bea
\eta &=& q-p \; , \\
\cos(2\theta) &=& \frac{1-p-q}{1-|q-p|} \, .
\eea
In formulae:
\be
G\left[\Xi_{q-p}\left(\Phi_1\left(U_{\frac12\arccos\left(\frac{1-p-q}{1-|q-p|}\right),\gr{\varphi}} \varrho_m 
U_{\frac12\arccos\left(\frac{1-p-q}{1-|q-p|}\right),\gr{\varphi}}^{\dag} \right)\right)\right] = 
T_{p,q} \gr{v}_m \; .
\ee
Clearly, all values of $p$ and $q$ may be reproduced by the open dynamics we considered 
by an appropriate choice of $\eta$ and $\theta$. $\square$\smallskip

It is apparent that the amplitude damping channel is needed to produce a bias between the two probabilities $p$ 
and $q$: for $p=q$, doubly stochastic maps are recovered without any such channel $(\eta=0)$, in agreement with the previous proposition. 

{\noindent} {\em The cellular automaton model.} 
Quantum cellular automata (discrete, translationally invariant, causal evolutions on a lattice) 
were first envisaged as computational models \cite{Margolus} 
and quantum simulators \cite{Feynmann}, although they have by now attracted attention 
both as hardware for specific quantum information processing tasks \cite{Macchiavello} and as models of causal quantum theories, 
including quantum field theory \cite{Ariano1,Ariano2} and quantum gravity \cite{Lloyd}. Such aims have focused on 
unitary automata, whose extension to more general CP-maps is little explored \cite{kennen}. Here, we consider a qubit lattice and employ our embedding of two-dimensional stochastic maps into dissipative qubit dynamics to define a class of noisy cellular automata that includes all probabilistic classical dynamics on the lattice, as well as more distinctly quantum dynamics.

Let us a consider a $1$-dimensional qubit lattice of length $N$, which is hosting an excitation transfer process 
(although our treatment can be extended to higher dimensions). 
We shall restrict to the single excitation subspace of the Hilbert space, spanned by the basis $\{\ket{n},1\le n\le N\}$,
where $\ket{n}$ represents the state with the $n^{\rm th}$ qubit in the excited state and all the other qubits in the ground state.
Let us now define the CP-map $\Omega^{(n)}_{\eta,\xi,\theta,\gr{\varphi}}$ as the map acting as the composition 
of a unitary $U^{(n)}_{\theta,\gr{\varphi}}$,
a dephasing $\Phi^{(n)}_{\xi}$ and an amplitude damping $\Xi^{(n)}_{\eta}$ on the two-dimensional subspace spanned by 
$\ket{n}$ and $\ket{n+1}$, and as the identity on the remainder of the single excitation subspace:
$
\Omega^{(n)}_{\eta,\xi,\theta,\gr{\varphi}} (\varrho) = \Xi^{(n)}_{\eta}\left(\Phi^{(n)}_{\xi}\left(U^{(n)}_{\theta,\gr{\varphi}}\varrho 
U^{(n)\dag}_{\theta,\gr{\varphi}}\right)\right),
$
where $\varrho$ is a density matrix with support in the single excitation subspace.
We can then define a noisy quantum cellular automaton on the lattice as the following map \cite{ArrighiQCA,ArrighiPQCA}
\be
\Omega_{\eta,\xi,\theta,\gr{\varphi}} = 
\bigotimes_{l\,{\rm even}}\Omega^{(l)}_{\eta,\xi,\theta,\gr{\varphi}} \bigotimes_{l\, {\rm odd}}\Omega^{(l)}_{\eta,\xi,\theta,\gr{\varphi}} \; .
\ee
Here, the ``odd'' and ``even'' prescriptions in the labels realise a partitioning of the lattice, taking into account the non-commutativity of CP-
maps acting on overlapping subsystems: One step of the automaton consists first in applying the map on disjoint pairs of neighbouring qubits, and then in applying the same operation shifted by one lattice position. 
Our lattice may be a ring -- in which case the map $\Omega^{N}_{\eta,\xi,\theta,\gr{\varphi}}$ 
acts on the state $\ket{N}$ and $\ket{1}$ -- or open -- in which case the map $\Omega^{N}_{\eta,\xi,\theta,\gr{\varphi}}$
is not applied \footnote{In the case of odd $N$, a lattice on a ring would require specific prescriptions for the partition used. 
We do not intend to enter such technicalities which have very little bearing on our findings.}.
Any unitary quantum cellular automaton may be realised, up to shift operations, by adopting such a partitioning,
based on the iteration of the same map between alternate pairs of neighbouring qubits \cite{arrighigrattage}. 
Although no corresponding general theorem exists for noisy CP-maps \cite{ArrighiProbQCA}, 
our maps are by construction causal and, on infinite lattices, invariant under the squared shift operator, 
so that it seems appropriate to maintain the denomination of (noisy) quantum cellular automata 
for them \cite{grossing}.

A classical transfer process may be modelled on such a lattice 
by a chain of identical stochastic transition matrices which, in the light of Proposition 2, 
can be represented by the CP-map $\Omega_{\eta,1,\theta,\gr{\varphi}}$ acting on diagonal (`classical') states, 
where the dephasing channel is set to a total dephasing, with strength $\xi=1$.
We have hence designed a class of cellular automata where one can study classical transfer by setting $\xi=1$, 
and then enter the quantum regime by decreasing the dephasing strength $\xi$ from $1$ to $0$. 
While still in a sense arbitrary, we argue our model is representative of the classical to quantum transition
in actual physical systems, in that it enacts such a transition entirely by changing a dephasing strength, which 
is the main decoherence mechanism in any open quantum system. In practical cases, dephasing results from the coupling  
with the environment, which sets the privileged basis.

Given $p$ and $q$ of a classical stochastic transfer process, one can construct the corresponding class of quantum cellular automata $\Omega_{\eta,\xi,\theta,\gr{\varphi}}$ by setting $\eta=p-q$ (whose sign will determine the privileged direction of travel of the excitation along the lattice) and $\theta=\arccos\left(\frac{1-p-q}{1-|q-p|}\right)/2$, and letting $\xi$ vary from the classical automaton for $\xi=1$ to the `most quantum' (where no dephasing acts and coherent off-diagonal terms are only suppressed by the amplitude damping) for $\xi=0$. The phases $\varphi_1$ and $\varphi_2$ are completely free, 
as one should expect since they cannot be determined by the limiting classical process where they do not appear at all. Such phases do potentially play a role in applications, as we will see shortly.

\noindent{\em Energy excitation transfer.}  We can now apply our model to the study of energy excitation transfer through the lattice \cite{Engel,lee,Hayes,Collini,Science,Mohseni2,rebentrost,rebentrost2,Plenio2,caruso,Olaya,
francesca,olayaview,Aspuru} 
by comparing, {\em at given local transition probabilities $p$ and $q$}, the performance of a classical process with that of quantum dynamics where 
coherent phases are allowed to develop and interfere along the chain. Note that the equality of the local transition probabilities ensures that all the difference between the classical and quantum cases is down to quantum coherence, in a very specific sense.

\begin{figure}[t!]
\begin{center}
\includegraphics[scale=0.44]{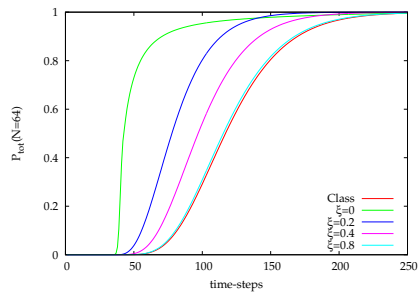}
\caption{Integrated probability of absorption through an open chain with $N=64$, $p=0.7$, $q=0.5$, 
$\varphi_1+\varphi_2=\pi$ (the probability only depends on the sum $\varphi_1+\varphi_2$) and various values of $\xi$, from classical ($\xi=1$, denoted by ``Class'') to most quantum ($\xi=0$).
 \label{bias}}
\end{center}
\end{figure}

We will study energy transfer by assuming the pure initial state $\ket{1}$, with a single excitation localised on the first site of the chain. 
The excitation absorption by a receptor located at site $\ket{N}$ and $\ket{N/2+1}$ for, respectively, an open chain and a ring of $N$ sites
(taking, for simplicity, $N$ to be even), will be modelled by a quantum measurement 
with elements $M_0=\id-\ket{N}\bra{N}$ and $M_1=\ket{N}\bra{N}$ (replacing $N$ with $N/2+1$ for a ring). 
The dynamics goes on until outcome $1$ occurs, whereby the excitation is captured at the receptor and the transfer process stops.
We are interested in the total probability of absorption after $t$ steps of the automaton (see Supplemental Material). 
As a preliminary investigation, we considered the optimal rate of measurement at the receptor site, in terms of maximising the 
absorption probability. 
In the classical case ($\xi=1$), where the only effect of measuring $0$ is renormalising the probability distribution, the optimal rate is 
measuring after every step of the automaton. In the quantum case, a failed absorption has the additional effect of destroying the off-diagonal 
terms involving the receptor site: nonetheless, it turns out that in the vast majority of cases we dealt with (see later) 
measuring after every step is still the optimal strategy, with very marginal gains when measuring every two steps in few specific cases. In the following, we will hence always consider absorption measurements performed at each step of the automaton.
Although analytical expressions are cumbersome even for the simplest configurations, 
the resulting conditional dynamics can be studied exactly for a very large number of sites. 

Let us start by considering an open chain of $N=64$ sites.
The case $p=0.7$ and $q=0.5$ is illustrated in Fig.~\ref{bias}. The advantage granted by quantum coherence is manifest, 
in that, for $\varphi_1+\varphi_2=\pi$, the presence of the off-diagonal terms of the density matrix increases dramatically the absorption probability at each step in the early dynamics (reflected in the increased slope in Fig.~\ref{bias}). 
Interestingly, after such an initial boost, systems with stronger quantum coherence are slower in saturating the integrated probability to $1$ than more classical counterparts.
Increasing the bias $\eta=p-q$ enhances the effect of the amplitude-damping channel, and thus diminishes the difference between the corresponding quantum and classical cases. 
In point of fact, note that, when $\eta=1$, the map 
becomes classical regardless of the choice of $\xi$ and other parameters: this is, so to speak, the ballistic limit, 
where the excitation is deterministically transferred through the chain in $N-1$ steps. 
While trivially optimal, this is not such an interesting regime when modelling stochastic transfer phenomena.
Further, notice that, because of the way we defined our partitioning, any case with $p=1$, including the one 
with $q=1$ that can be obtained by chains of unitary swap operations, also results in ballistic transfer.
The benefit granted by stronger amplitude damping in our model is distinct from the seminal cases of noise assistance  
flagged up in \cite{Mohseni2,rebentrost,rebentrost2,Plenio2,caruso}, where local dephasing is responsible 
for suppressing destructive interference 
\footnote{Also note that, although expedient to comply with the 
literature on single-qubit channels, the terminology ``amplitude damping'' is in our case slightly misleading, since here the 
environment is not locally draining excitations, but rather acting at the interface between two qubits 
by pushing the excitation along a privileged direction.}. As we will see, the latter can also be reproduced within our framework.

\begin{figure}[t!]
\begin{center}
\subfigure{\includegraphics[scale=0.335]{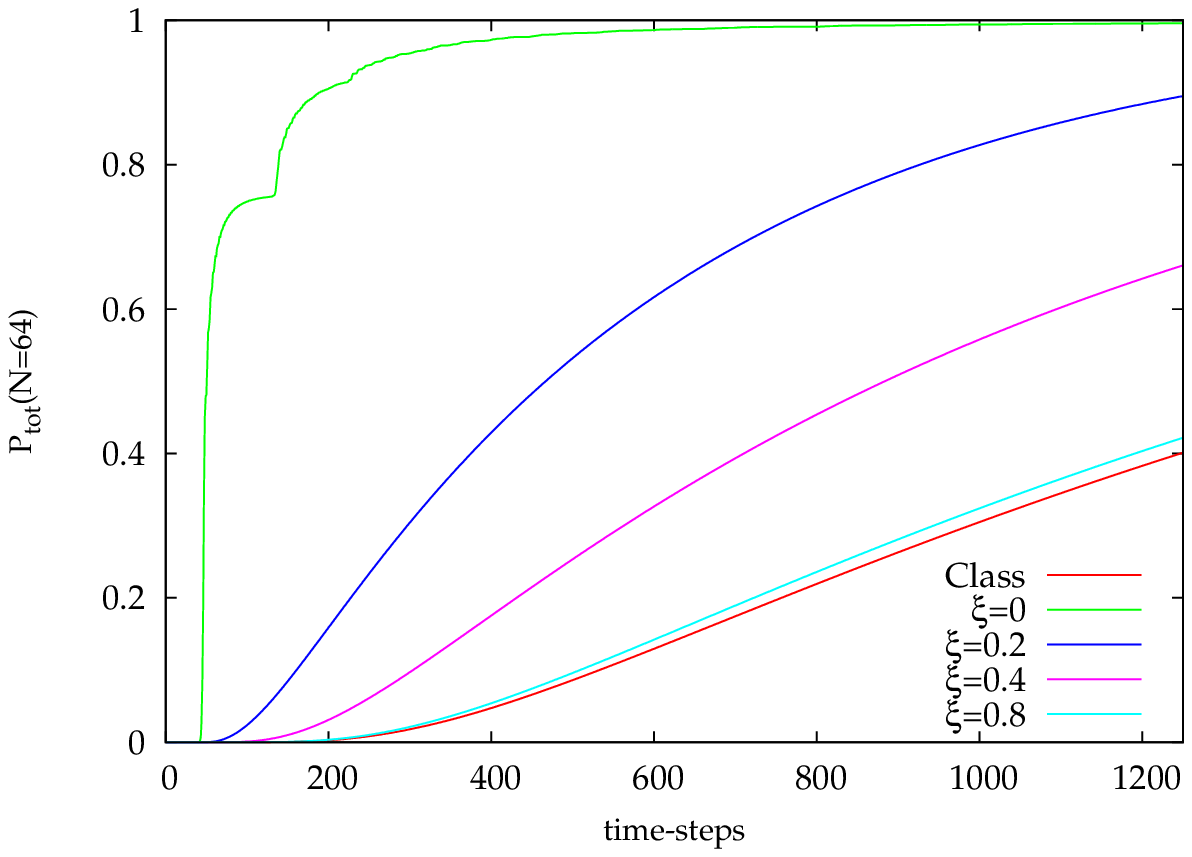}}
\subfigure{\includegraphics[scale=0.335]{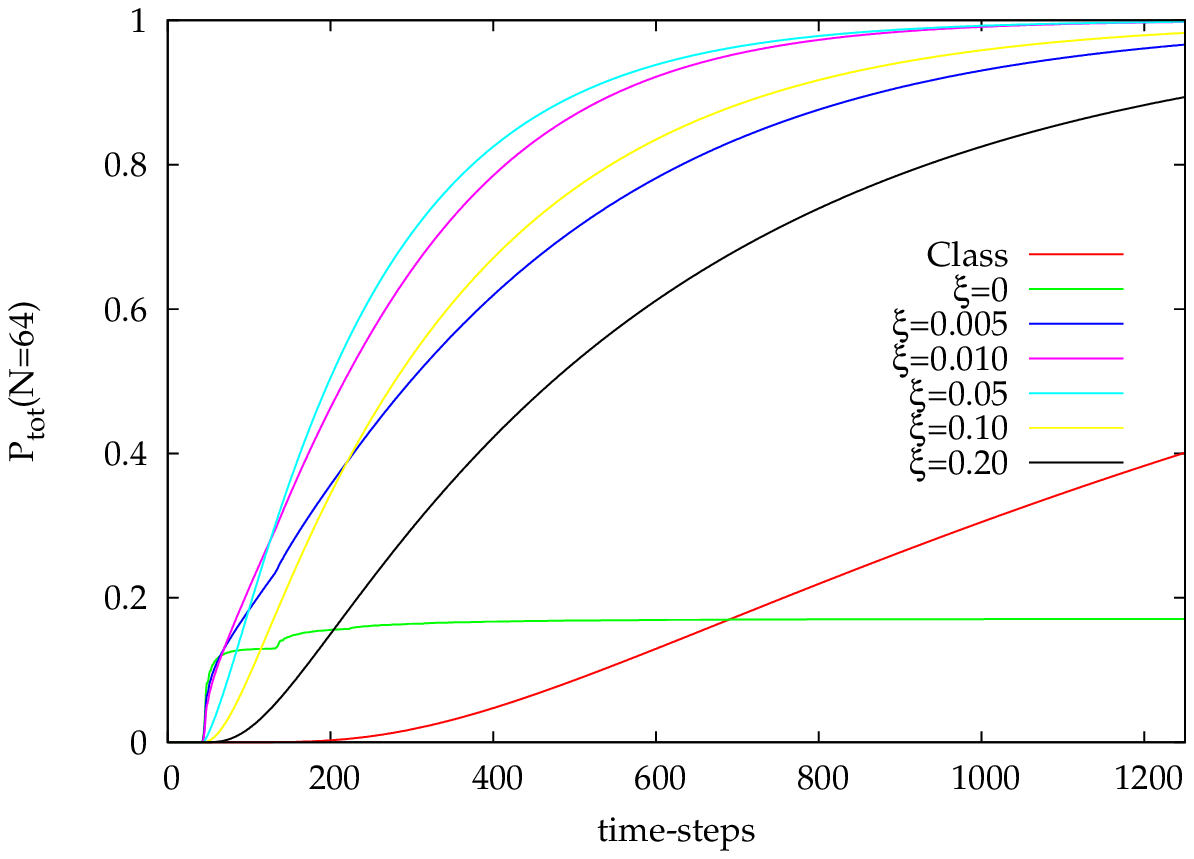}}
\caption{Integrated probability of absorption through an open chain with $N=64$, $p=0.5$, $q=0.5$, 
various values of $\xi$, from classical ($\xi=1$, denoted by ``Class'') to most quantum ($\xi=0$), 
and phases $\varphi_1+\varphi_2=\pi$ (a) and $\varphi_1=\varphi_2=0$ (b). \label{nobias}}
\end{center}
\end{figure}

Intriguing effects become apparent setting $p=q=0.5$, as reported in Fig.~\ref{nobias}(a) for an open chain with optimised 
phases (such that $\varphi_1+\varphi_2=\pi$). 
In this instance, the gap between quantum and classical dynamics is at its widest, and purely quantum distinctive features 
emerge. In particular, the integrated absorption probability shows stationary points, 
whereby the instantaneous absorption probability is zero, followed by sudden increases. 
This effect, for which we provide heuristic analytical evidence in the Supplemental Material, is a manifestation of destructive interference due to the off-diagonal terms of the density matrix, and 
disappears as soon as any amount of dephasing is introduced. It is however a purely quantum effect, which could in principle 
be observed. 

Dephasing-assisted transfer is apparent in Fig.~\ref{nobias}(b), where we set parameters as in \ref{bias}(a) except for the 
coherent phases $\gr{\varphi}$ (also, we scan a different set of values for $\xi$). In this case destructive interference is clear in the quantum case ($\xi=0$), where the integrated probability soon encounters a plateau, and is suppressed as soon as some dephasing noise is introduced. An optimal value around $\xi=0.05$ can be determined with this choice of parameters. 
This is at variance with the stationarity encountered for optimised phases, 
where each stationary point is followed by a steep ramp of constructive interference.
Our framework is hence capable of highlighting the dependence of noise-assistance 
on the phases of coherent interactions (the unitary $U$, in our discrete treatment): for certain choices of phases (such as the one 
in Fig.\ref{nobias}(a)), dephasing noise helps only marginally and at long times (after the initial quantum boost) \footnote{Also note 
that the action of local dephasing on each qubit is equivalent to the action of our dephasing channel in the 
single excitation subspace.}.

\begin{figure}[t!]
\begin{center}
\subfigure[]{\includegraphics[scale=0.33]{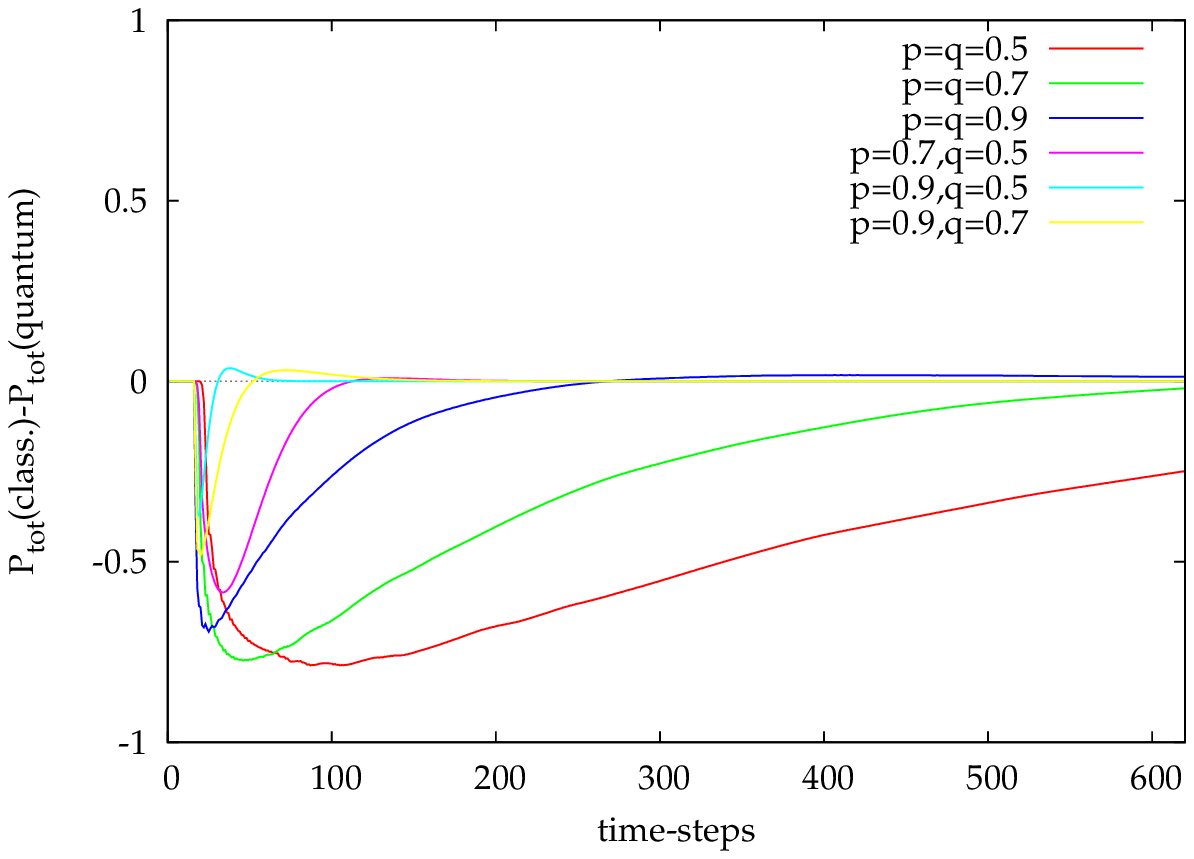}}
\subfigure[]{\includegraphics[scale=0.33]{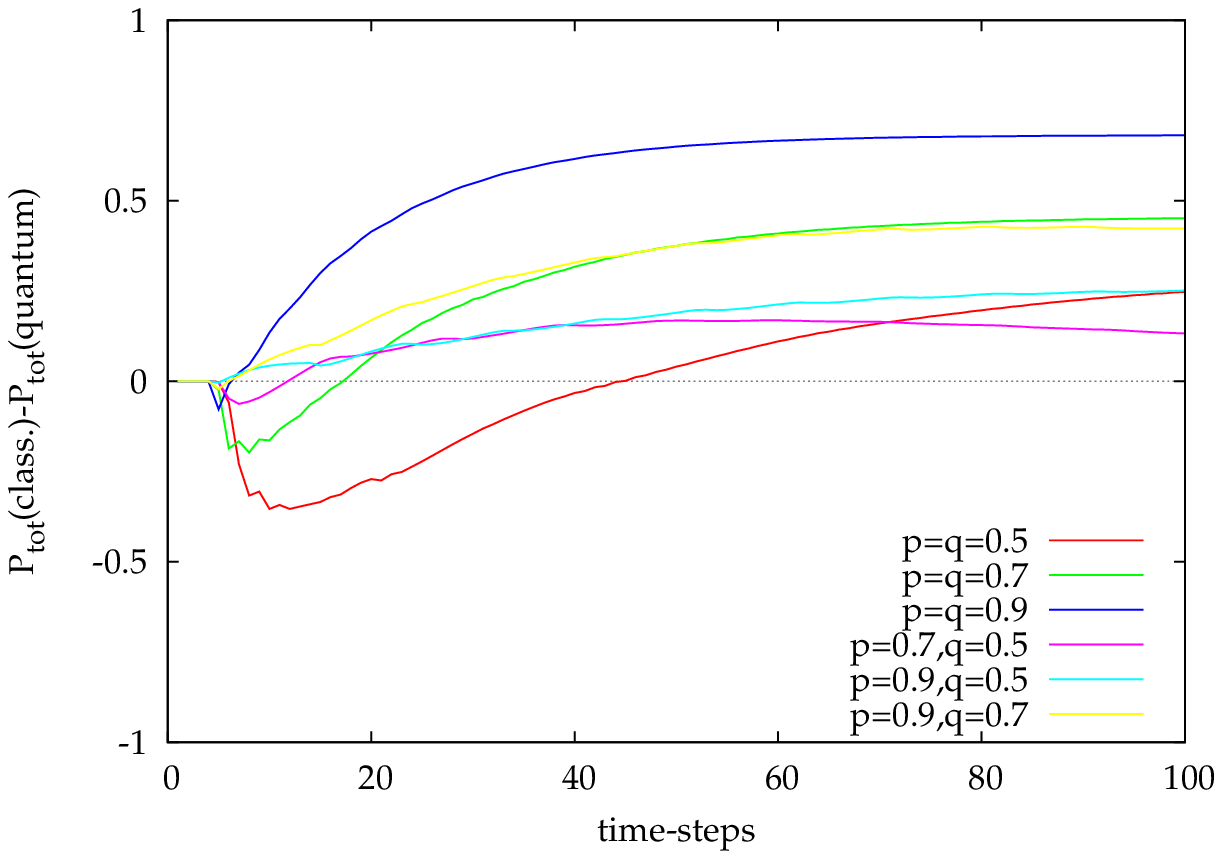}}
\caption{Difference between most classical ($\xi=1$) and most quantum ($\xi=0$) integrated probabilities of absorption 
through a ring with $\varphi_1=\varphi_2=0$,  
various values of $p$ and $q$, and $N=64$ (a) and $N=18$ (b). \label{ring}}
\end{center}
\end{figure}

The case of a ring with $N=64$ is depicted in Fig.~\ref{ring}(a), and confirms that the advantage granted by quantum coherence 
is most apparent in the case $p=q=0.5$, and tends to vanish as the difference between $p$ and $q$ increases. 
The quantum advantage in the transfer probability critically depends on the number of qubits $N$: with more qubits, 
the effect of constructive interference becomes more relevant and enduring. 
This advantage is reminiscent of the speed-up occurring in random \cite{Aharonov} 
or Hamiltonian \cite{Childs} quantum walks -- where by `random' quantum walk we refer to dynamics featuring a 
coin Hilbert space -- which share similarities with our approach.
For instance, the analogous of Fig.~\ref{ring}(a) for $N=18$ is reported in Fig.~\ref{ring}(b), and shows that the classical maps surpasses their quantum counterparts (with optimally chosen phases) 
after at most $70$ time-steps. 

To summarise, we have introduced a class of causal CP-maps on a qubit lattice, which generalise the notion of unitary cellular automaton
and embed all possible stochastic 
maps on classical probability distributions on the lattice, and then applied such a class of noisy automata to contrast 
the performance and behaviour of classical and quantum excitation transfer processes. Our discrete model 
is capable of highlighting coherent effects, such as noise-assistance, can be applied to very large systems and 
allows one to treat conditional quantum dynamics exactly. 
Besides, the application of our framework may be extended to more general communication problems, 
such as quantum and classical capacities \cite{carusotto} or transfer fidelities along the chain \cite{Bose,Albanese,Cristo,daniel1,vittorio,daniel2}.

\medskip
\noindent{\em We acknowledge joint financial support from Mischa Stocklin and UCL through the 
Impact scholarships scheme, as well as discussions with M. Stocklin, A. Olaya-Castro, F. Fassioli, F. Brandao, S. Virmani, V. Giovannetti and P. Arrighi.}


\hfill
\newpage

\appendix

\section{Supplemental Material}

\subsection{Total probability of absorption}
The quantity we keep track of during the evolution of the automaton is the total
probability of absorption after $t$ time steps of the dynamics. More precisely, this is the total integrated probability 
that after $t$ iterations of the automaton after which a measurement on site $N$ has been performed at every time step, 
the system has measured the excitation on the last site $N$. In formulae:
\begin{small}
\begin{eqnarray}
P_N^{tot}(t=1) & = & \varrho_{NN}(t=1) \nonumber \\
P_N^{tot}(t=2) & = & [1-\underbrace{(1-\varrho_{NN}(t=1))}_{1-P_N^{tot}(t=1)}\underbrace{(1-\varrho_{NN}(t=2))}_{1-P_N^{tot}(t=2)}] \nonumber \\
P_N^{tot}(t=3) & = & [1-\underbrace{(1-\varrho_{NN}(t=1))}_{1-P_N^{tot}(t=1)}\underbrace{(1-\varrho_{NN}(t=2))}_{1-P_N^{tot}(t=2)} \nonumber \\
 & & \cdot \underbrace{(1-\varrho_{NN}(t=3))}_{1-P_N^{tot}(t=3)}] \nonumber \\
\dots & &  
\end{eqnarray}
\end{small}
and so on.

\subsection{Coherent effects on absorption}

Recalling the partitioning induced on the lattice by the automaton, the global operators
composing the map (eq.($11$)-main text) are block diagonal matrices, where for the \textit{even}
partitioning all the blocks are $2\times 2$, while for the \textit{odd} partitioning the 
first and last blocks are just the $1$-dimensional identity. Because of the neighbouring scheme
of the automaton, only $3$ sites directly interact with each other at each iteration. 

\begin{figure}[t!]
\begin{center}
\includegraphics[scale=0.6]{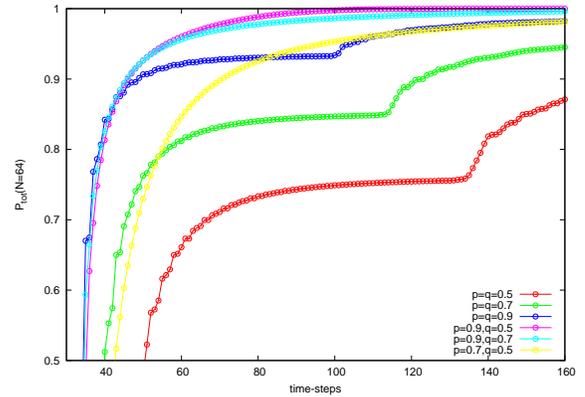}
\caption{Integrated probability of absorption through an open chain with $N=64$, $\varphi_1+\varphi_2=\pi$ and various values of $p,q$. The dephasing channel is switched off ($\xi=0$).  \label{fig:nodeph}}
\end{center}
\end{figure}

\begin{figure}[t!]
\begin{center}
\includegraphics[scale=0.6]{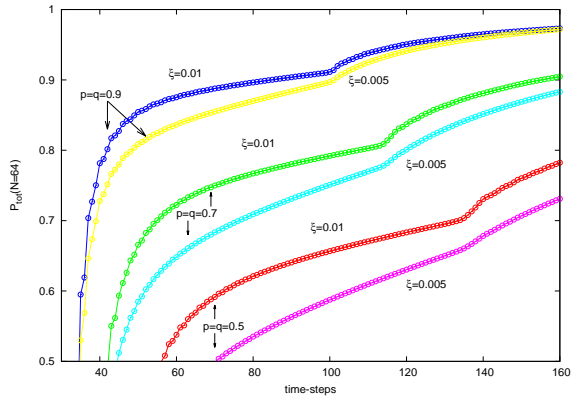}
\caption{Integrated probability of absorption through an open chain with $N=64$, $\varphi_1+\varphi_2=\pi$ and different values of $p,q$. The dephasing channel is now switched on: $pd\equiv \xi=(0.005;0.01)$.  \label{fig:deph}}
\end{center}
\end{figure}

Focusing only on the last $3$ sites of the $N\times N$ global density matrix $\varrho$, the action of
the {\em unitary part} of the map (eq.($11$)-main text) is given by \footnote{Let us remind the reader that all
the entries in the $N$th row and column of the global state of the system $\varrho$ are set to $0$ due
to the measurement that took place at the previous time step.}:  
\begin{widetext}
\begin{small}
\be
U_{(3\times 3)}^{even}U_{(3\times 3)}^{odd}\varrho_{(3\times 3)}U_{(3\times 3)}^{even\dagger}U_{(3\times 3)}^{odd  \dagger},\quad\mbox{ with} \qquad \varrho_{(3\times 3)}= 
\begin{bmatrix}
\varrho_{N-2,N-2} & \varrho_{N-2,N-1} & 0 \\
\varrho_{N-2,N-1}^* & \varrho_{N-1,N-1} & 0 \\
0 & 0 & 0 \\
\end{bmatrix}, \quad U_{(3\times 3)}^{even}=
\begin{bmatrix}
u_{22} & 0 & 0 \\
0 & u_{11} & u_{12} \\
0 & u_{21} & u_{22} \\
\end{bmatrix}, \quad U_{(3\times 3)}^{odd}=
\begin{bmatrix}
u_{11} & u_{12} & 0 \\
u_{21} & u_{22} & 0 \\
0 & 0 & 1 \\
\end{bmatrix},
\ee
\end{small}
\end{widetext}
which gives a probability of absorption at time $t+1$:
\begin{small}
\be
\varrho_{NN}^{(t+1)}=|u_{21}|^2\left[ |u_{21}|^2\varrho_{N-2,N-2}^{(t)}+|u_{22}|^2\varrho_{N-1,N-1}^{(t)}+\left(
u_{21}u_{22}^*\varrho_{N-2,N-1}^{(t)}+c.c \right)\right] \, .
\ee
\end{small}
In the case $p=q=0.5$, $\varphi_1=\pi$, $\varphi_2=0$ (the probability of absorption only depends on the sum $\varphi_1+\varphi_2$), the equation above reads 
\begin{small}
\be
\varrho_{NN}^{(t+1)}=\frac12\left[ \frac12 \varrho_{N-2,N-2}^{(t)}+\frac12\varrho_{N-1,N-1}^{(t)}-
\varrho_{N-2,N-1}^{(t)}\right] \,
\ee
\end{small}
(and all the entries of $\varrho$ are real). 
The stationarity of the total probability of absorption we can see in Fig.~\ref{fig:nodeph} 
occurs when $\varrho_{NN}^{(t+1)} \simeq 0$ at some time-step. 
It can be easily shown that such a condition is only possible when 
$\varrho_{N-2,N-2}^{(t)}\simeq\varrho_{N-1,N-1}^{(t)}$ and when the off-diagonal term 
$\varrho_{N-2,N-1}^{(t)}=\sqrt{\varrho_{N-1,N-1}^{(t)}\varrho_{N-2,N-2}^{(t)}}$.
By pinching ~\cite{pincio}, it can be seen that this is the maximum value of $\varrho_{N-2,N-1}^{(t)}$ 
compatible with the positivity condition on $\varrho$. This is the reason why any amount of 
dephasing will prevent the cancellation above, and hence stationarity from happening (Fig.~\ref{fig:deph}).

Recalling that the amplitude damping channel (eq.($3$)-main text) also suppresses coherent off-diagonal terms 
by a factor $\sqrt{1-|q-p|}$, the same argument heuristically explains why stationarity is ruled out for
any dynamics with $p\neq q$ (Fig.~\ref{fig:nodeph}). 

\end{document}